\documentclass[aps,prb,twocolumn,superscriptaddress,showpacs,preprintnumbers,amsmath,amssymb]{revtex4}
\usepackage{amsmath}
\usepackage{dsfont}
\usepackage{amstext}

\usepackage{amssymb}
\usepackage{amsbsy}
\usepackage{amsthm}
\usepackage{epsfig}
\usepackage{graphicx}
\usepackage{bbm}
\usepackage{hyperref}
\usepackage{color}

\newcommand{\Ref}[1]{Ref.~\onlinecite{#1}}

\newcommand{\bse}{{\boldsymbol{e}}}

\newcommand{\ie}{{\emph{i.e.~}}}
\makeatletter

\newcommand{\Rmnum}[1]{\expandafter\@slowromancap\romannumeral #1@}
\makeatother
\newcommand{\imth}{\hspace{1pt}\mathrm{i}\hspace{1pt}}

\newcommand{\eg}{{\emph{e.g.~}}}

\newcommand{\mbz}{{\mathbb{Z}}}
\newcommand{\dof}{\emph{d.o.f.}~}
\newcommand{\bea}{\begin{eqnarray}}
\newcommand{\eea}{\end{eqnarray}}

\begin{document}
\title{Quantum phase transitions between bosonic symmetry protected topological phases in two dimensions: emergent $QED_3$ and anyon superfluid}

\author{Yuan-Ming Lu}
\affiliation{Department of Physics, University of California, Berkeley, CA 94720, USA}
\affiliation{Materials Science Division, Lawrence Berkeley National Laboratories, Berkeley, CA 94720}

\author{Dung-Hai Lee}
\affiliation{Department of Physics, University of California, Berkeley, CA 94720, USA}
\affiliation{Materials Science Division, Lawrence Berkeley National Laboratories, Berkeley, CA 94720}

\begin{abstract}
Inspired by Chern-Simons effective theory description of symmetry protected topological (SPT) phases in two dimensions, we present a projective construction for many-body wavefunctions of SPT phases. Using this projective construction we can systematically write down trial wavefunctions of SPT phases on a lattice. An explicit example of SPT phase with $U(1)$ symmetry is constructed for two types of bosons with filling $\nu_{b_1}=\nu_{b_2}=\frac12$ per site on square lattice. We study continuous phase transitions between different $U(1)$-SPT phases based on projective construction. The effective theory around the critical point is emergent $QED_3$ with fermion number $N_f=2$. Such a continuous phase transition however needs fine tuning, and in general there are intermediate phases between different $U(1)$-SPT phases. We show that such an intermediate phase has the same response as an anyon superconductor, and hence dub it ``\emph{anyon superfluid}". A schematic phase diagram of interacting bosons with $U(1)$ symmetry is depicted.
\end{abstract}

\maketitle

\section{Introduction}

Topological insulators and superconductors\cite{Hasan2010,Hasan2011,Qi2011}, which has drawn lots of interest recently, belong to a large class of disordered gapped phases dubbed symmetry protected topological (SPT) phases\cite{Chen2013}. One definitive feature of a SPT phase, is the existence of gapless boundary excitations which is protected by certain symmetries. In the absence of any symmetry, without gap closing these states can always be continuously connected to a featureless gapped state (called the trivial phase) which is a direct product of local degrees of freedom. When symmetry is preserved, however, these SPT phases are separated from the trivial phase by a phase transition. Extensive studies of fermonic SPT phases have been done based on free fermion band structures, including the classification of non-interacting fermionic SPT phases with various symmetries\cite{Schnyder2008,Kitaev2009}. On the other hand, bosonic SPT phases require strong interaction to realize and are much less well understood\cite{Chen2013,Chen2011b,Levin2012,Levin2012a,Lu2012a,Senthil2013,Vishwanath2013,Xu2013,Liu2013,Chen2012}, not to mention the phase transitions between them. Since SPT phases are gapped phases which preserves the symmetry, the phase transitions between different SPT phases are beyond the description of Landau's symmetry breaking theory\cite{Landau1937,Landau1937a}. Can there be any non-Landau-Ginzburg-Wilson type\cite{Senthil2004,Senthil2004a} continuous quantum phase transition\cite{Sachdev2011} between two different bosonic SPT phases? If not, what are the intermediate phases between two different bosonic SPT phases? In this work we'll address these questions, focusing on bosonic SPT phases protected by $U(1)$ symmetry in two dimensions.

\begin{figure}
 \includegraphics[width=0.5\textwidth]{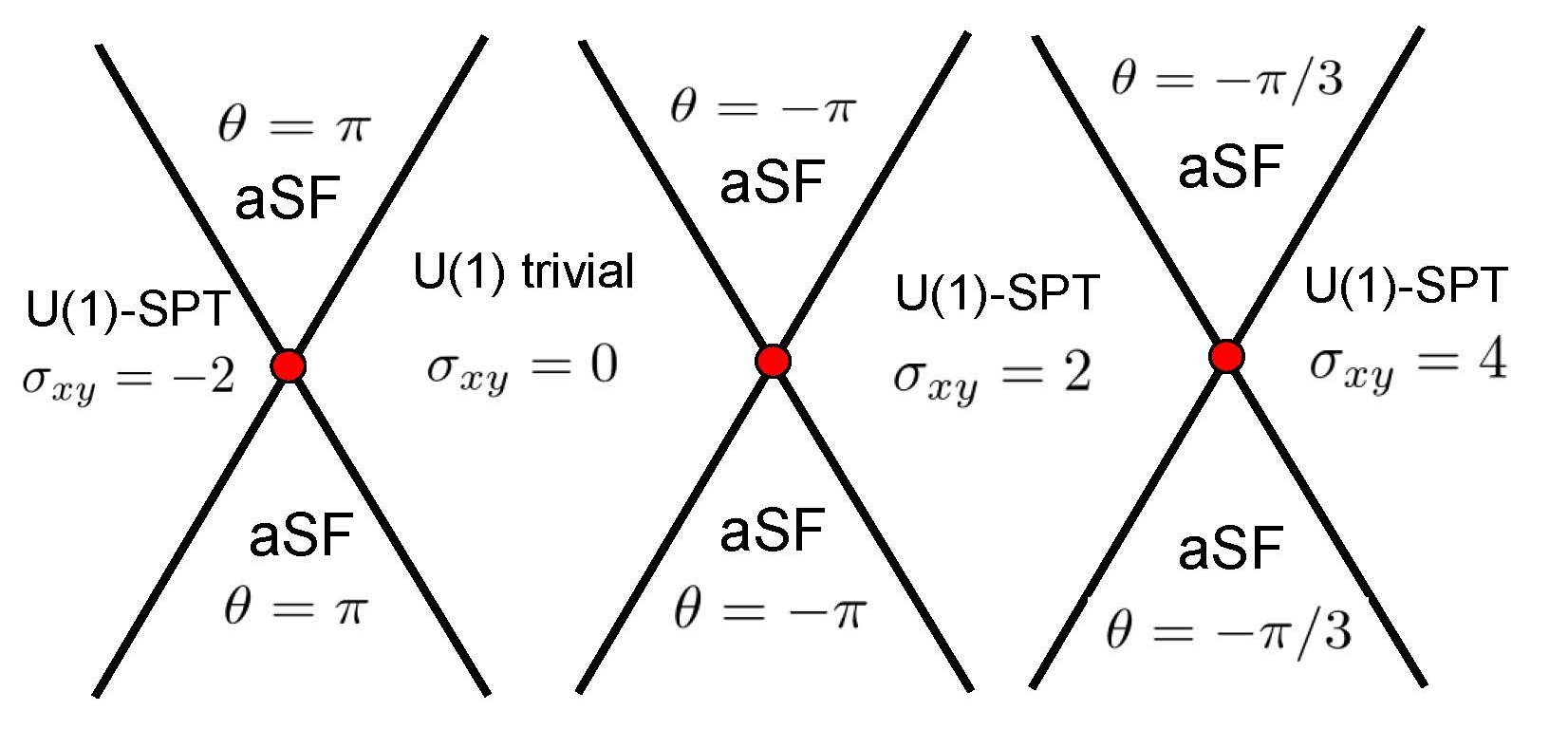}
\caption{(color online) A schematic phase diagram of interacting bosons with $U(1)$ symmetry (associated with boson charge conservation) in two dimensions. It includes trivial boson insulator, bosonic $U(1)$-SPT insulators and gapless anyon superfluid (aSF) phases. Different bosonic insulators are featured by their Hall conductance $\sigma_{xy}=2q,~q\in\mbz$, in the presence of $U(1)$ symmetry associated with boson number conservation. An aSF phase spontaneously breaks $U(1)$ symmetry and is featured by superfluid response and a \emph{quantized} Chern-Simons term in (\ref{response:anyon SF}). $\theta=\pi/(1-2q)$ denotes the statistical angle of anyon, in the aSF phase between two bosonic $U(1)$-symmetric insulators with $\sigma_{xy}=2q$ and $\sigma_{xy}=2q-2$. Solid lines denote phase boundaries between $U(1)$-SPT phases and anyon superfluids, which are connected by a continuous phase transition with effective theory (\ref{aSF-SPT}). Each red circle denotes a tricritical point, whose effective theory is described by emergent $QED_3$ with fermion number $N_f=2$.}\label{fig:schematic phase diagram}
\end{figure}

In the presence of $U(1)$ symmetry related to boson charge conservation, there is an infinite number of distinct gapped boson phases\cite{Chen2013} labeled by an integer $q\in\mbz$ in two space dimensions. Each gapped bosonic phase is featured by its quantized Hall conductance\cite{Lu2012a,Levin2012a,Senthil2013,Liu2013,Chen2012} $\sigma_{xy}=2q$ in unit of $e_b^2/h$ ($e_b$ is the unit $U(1)$ charge carried by bosons). The trivial gapped boson phase corresponds to $q=0$. For any $q\neq0$ phase  there is either gapless excitations at the boundary or the symmetry is spontaneously broken. These $U(1)$-SPT phases can be described by a $U(1)\times U(1)$ Chern-Simons theory\cite{Lu2012a,Levin2012a,Senthil2013}, where the physical $U(1)$ symmetry is implemented by a charge vector\cite{Wen1992}. Based on such effective Chern-Simons theory\cite{Lu2012a} we develop a projective construction\cite{Wen1999} of the ground state wavefunctions for bosonic $U(1)$-SPT phases in $2+1$-D. This construction not only provides a systematic way to write down trial wavefunctions for bosonic SPT phases  on lattices, but also enable us to study continuous quantum phase transitions between different U(1)-SPT phases. As a byproduct it also provides a systematic way to write down trial wavefunctions of bosonic SPT phases on a lattice. For continuous transitions where the Hall conductance change by two units, we show the critical theory is $QED_3$ with $N_f=2$, \ie two flavors of massless Dirac fermions couple to a \emph{noncompact} $U(1)$ gauge field, where the microscopic $U(1)$ symmetry leads to conservation of $U(1)$ gauge flux at the critical point. Such a critical theory has been studied in the context of algebraic spin liquid\cite{Affleck1988a,Marston1989,Kim1999,Rantner2001,Franz2001,Hermele2004,Hermele2005,Hermele2005} and is known to describe an interacting conformal fixed point\cite{Rantner2001} (beyond free quasiparticle descriptions). However, this $\Delta\sigma_{xy}=2$ transition generically will break into two consequently $\Delta\sigma_{xy}=1$ transitions in the absence of extra symmetries. We show that the intermediate phase is a \emph{anyon superfluid} (aSF)\cite{CHEN1989,Fetter1989,Fradkin1990} with spontaneous $U(1)$ symmetry breaking. Based on these results, a schematic phase diagram of interacting bosons with $U(1)$ symmetry is shown in FIG. \ref{fig:schematic phase diagram}.

The outline of the paper is as follows. In section \ref{CHERN SIMONS APPROACH} we briefly review the Chern-Simons effective theory of bosonic SPT phases in $2+1$-D, with an emphasis on $U(1)$-symmetric SPT phases. In section \ref{PROJECTIVE CONSTRUCTION} we develop a projective construction of the ground state wavefunctions for bosonic $U(1)$-SPT phases, where an example of two species of half-filled bosons on square lattice is studied in detail. In section \ref{QUANTUM PHASE TRANSITION} we study the continuous phase transitions between two different $U(1)$-SPT phases and derive the critical theory. We show that generically there is an intermediate phase between two adjacent $U(1)$-SPT phases. We conclude with some remarks in section \ref{CONCLUSION}.

\section{Chern-Simons approach to symmetry protected topological phases: a brief review}\label{CHERN SIMONS APPROACH}

In this section we briefly review the Chern-Simons approach to bosonic/fermionic symmetry protected topological (SPT) phases\cite{Lu2012a}. The low-energy effective theory of SPT phases is manifest in this approach, allowing us to study the quantum phase transition between different SPT phases.

The Chern-Simons approach was firstly introduced\cite{Read1990,Blok1990a,Wen1992,Frohlich1991} to classify and characterize Abelian fractional quantum Hall (AFQH) states in two dimensions. To be specific, the long-wavelength effective field theory describing a generic (multicomponent) AFQH fluid is written in terms of compact $U(1)$ gauge fields $\{a_\mu^I\}$:
\begin{eqnarray}\label{bare CS}
\mathcal{L}_{CS}=-\frac{1}{4\pi}\sum_{I,J=1}^N\epsilon^{\mu\nu\lambda}a^I_\mu {\bf K}_{I,J}\partial_\nu a^J_\lambda
\end{eqnarray}
where ${\bf K}_{I,J}$ is a symmetric $N\times N$ matrix with integer entries (summing over repeated indices $\mu,\nu,\lambda=t,x,y$ is always assumed in the paper). There are $N$ different types of conserved ``electron"\footnote{Here the word ``electrons" have a different meaning than the usual one: we call every microscopic particle (physical degree of freedom) in the system an ``electron". An ``electron" in our context can be either a boson or a fermion.} currents $\{J_I^\mu|1\leq I\leq N\}$ in such a AFQH liquid, given by
\begin{eqnarray}\label{duality:electron current}
J_I^\mu=\frac1{2\pi}\epsilon^{\mu\nu\lambda}\partial_\nu a_\lambda^I.
\end{eqnarray}
AFQH liquids can be viewed as a condensate of ``composite bosons"\cite{Zhang1989,Read1989,Wen1992} with various type of vortex (quasiparticle) excitations. (\ref{duality:electron current}) is the expression of composite boson current after the standard non-relativistic duality transformation\cite{Fisher1989,Wen1992}. Each composite boson is a composite of electrons and fluxes\cite{Wilczek1982,Wilczek1982a}) and hence has the same density/current as electrons. In the dual theory the vortex (or quasiparticle) current $j^\mu_I$ (of the $I$-th type) couples minimally to $U(1)$ gauge field $a_\mu^I$. In such a Chern-Simons theory the coupling of ``electrons" (or composite bosons) to the physical electromagnetic gauge field is specified by a \emph{charge vector}\cite{Wen1992} ${\bf t}=(t_1,t_2,\cdots,t_N)^T$. Including these features the low-energy effective theory for a generic AFQH liquid (we set $e=\hbar=1$ in most of the paper) is given by:
\begin{eqnarray}\label{bulk lagrangian}
\mathcal{L}_{eff}=\mathcal{L}_{CS}-\sum_I a_\mu^Ij_I^\mu-\frac{\epsilon^{\mu\nu\lambda}}{2\pi}\sum_It_IA_\mu\partial_\nu a^I_\lambda.
\end{eqnarray}
Writing $j_\mu^I=l^I j_\mu$ and integrating out $a_\mu^I$ gauge fields one immediately obtains all topological features of this AFQH liquid:
\begin{eqnarray}
&\notag\mathcal{L}^\prime_{eff}=\frac{\epsilon^{\mu\nu\lambda}}{4\pi}({\bf t}^T{\bf K}^{-1}{\bf t})A^\mu\partial_\nu A_\lambda\\
&\label{response Lagrangian}+({\bf l}^T{\bf K}^{-1}{\bf t})A_\mu j^\mu\\
&\notag+{\pi}({\bf l}^T{\bf K}^{-1}{\bf l})\epsilon_{\mu\nu\lambda}j^\mu\frac{\partial_\nu}{\square}j^\lambda.
\end{eqnarray}
Here ${\bf l}=(l^1,\cdots,l^N)^T$ is an integer\footnote{The requirement that a vector ${\bf l}$ must be an integer vector comes from the fact that $a_\mu^I$ are all compact $U(1)$ gauge fields.} vector characterizing a quasiparticle in AFQH liquid, which is a conglomerate of $l^I$ vortices of $I$-th type. In the following we'll simply call it a quasiparticle ${\bf l}$. $\square$ represents the Laplace operator in $2+1$-D. The 1st term in (\ref{response Lagrangian}) is the Hall response of the AFQH liquid
\begin{eqnarray}
\sigma_{xy}=\frac{e^2}{h}{\bf t}^T{\bf K}^{-1}{\bf t}.
\end{eqnarray}
The 2nd term describes the electric charge of each quasiparticle ${\bf l}$:
\begin{eqnarray}
Q_{\bf l}=-{\bf l}^T{\bf K}^{-1}{\bf t}.
\end{eqnarray}
The 3rd term is a Hopf Lagrangian\cite{Wilczek1983} describing the self exchange statistics of quasiparticle ${\bf l}$. Its statistical angle is
\begin{eqnarray}\label{statistics:self}
\theta_{\bf l}=\pi{\bf l}^T{\bf K}^{-1}{\bf l}\mod2\pi.
\end{eqnarray}
which is $0$ for bosons, $\pi$ for fermions and otherwise for Abelian anyons. One can further show the mutual (braiding) statistical angle for two quasiparticles ${\bf l}$ and ${\bf l}^\prime$ is
\begin{eqnarray}\label{statistics:mutual}
\theta_{{\bf l}\odot{\bf l}^\prime}=2\pi{\bf l}^T{\bf K}^{-1}{\bf l}^\prime\mod2\pi.
\end{eqnarray}
On an open manifold the gauge invariance of Lagrangian (\ref{bare CS}) implies the existence of gapless edge excitations. Effective edge theory (\eg along $\hat{x}$ direction) associated with bulk theory (\ref{bulk lagrangian}) is given by\cite{Wen1995}
\begin{eqnarray}
&\notag\mathcal{L}_{edge}=\frac1{4\pi}\sum_{I,J}\big({\bf K}_{I,J}\partial_t\phi_I\partial_x\phi_J-{\bf V}_{I,J}\partial_x\phi_I\partial_x\phi_J\big)\\
&-\frac{1}{2\pi}\sum_It_I\big(A_0\partial_x\phi_I-A_x\partial_t\phi_I\big).\label{edge lagrangian}
\end{eqnarray}
where $A_{0,x}$ are the external $U(1)$ electromagnetic gauge fields and $\{\phi_I\}$ are chiral boson fields. ${\bf V}_{I,J}$ is a positive-definite real symmetric matrix.

An important feature of the topological order\cite{Wen1990a} (or long range entanglement\cite{Kitaev2006a,Levin2006,Li2008}) in a FQH liquid is its ground state degeneracy (GSD) on a closed 2-manifold\cite{Wen1990b}. For example the GSD of the AFQH liquid described by effective theory (\ref{bare CS}) is\cite{Keski-Vakkuri1993}
\begin{eqnarray}
GSD_g=|\det{\bf K}|^g.
\end{eqnarray}
on a Riemann surface of genus $g$. One immediately notices that when $\det{\bf K}=\pm1$, the ground state described by (\ref{bulk lagrangian}) is always non-degenerate on any closed manifold. Moreover from (\ref{statistics:self}) and (\ref{statistics:mutual}) one can see that all the quasiparticles (characterized by an integer vector ${\bf l}$) are either bosons or fermions with bosonic (trivial) mutual statistics. In this case the corresponding $2+1$-D gapped phase has no long range entanglement and no quasiparticles with fractional statistics, which are basic features of symmetry protected topological (SPT) phases in $2+1$-D. Besides the edge excitations of a SPT phase contain an equal number of right and left movers, so that in the absence of symmetry protection they can be all gapped out. From the edge effective theory (\ref{edge lagrangian}) it's straightforward to show that the number of right and left movers $(n_+,n_-)$ on the edge is nothing but the \emph{signature} of matrix ${\bf K}$. In the simplest case of a $1\times1$ matrix ${\bf K}=K\in\mathbb{Z}$ the chirality of the edge mode is determined by the sign of $K$. As a result if effective theory (\ref{bulk lagrangian}) describes a SPT phase in $2+1$-D, it must satisfy
\begin{eqnarray}
\det{\bf K}=(-1)^{\text{dim}({\bf K})/2}.
\end{eqnarray}
so the edge states have equal number of left and right movers.

When $\det{\bf K}=\pm1$, it's easily seen that if at least one diagonal elements of ${\bf K}$ matrix are \emph{odd} integers, there are \emph{fermionic} excitations ($\exists{\bf l}~s.t.~\theta_{\bf l}=\pi$) and effective theory (\ref{bulk lagrangian}) describes a short range entangled (SRE) \emph{fermionic} state in $2+1$-D. If all diagonal elements of ${\bf K}$ matrix is \emph{even}, on the other hand, all the quasiparticles are bosonic ($\theta_{\bf l}=0,~\forall~{\bf l}$) hence the effective theory (\ref{bulk lagrangian}) describes a SRE \emph{bosonic} state in $2+1$-D. In \Ref{Lu2012a} it was shown that many $2+1$-D SPT phases (with various unitary/anti-unitary symmetries) can be described using a $2\times2$ ${\bf K}$ matrix, \ie
\begin{eqnarray}\label{K mat:boson}
{\bf K}=\begin{pmatrix}0&1\\1&0\end{pmatrix}
\end{eqnarray}
for \emph{bosonic} SPT phases and
\begin{eqnarray}\label{K mat:fermion}
{\bf K}=\begin{pmatrix}1&0\\0&-1\end{pmatrix}
\end{eqnarray}
for \emph{fermionic} SPT phases.

The key step from the Chern-Simons effective theory (\ref{bulk lagrangian}) to classification and characterization of $2+1$-D SPT phases is to incorporate symmetry into the effective theory. In the study of AFQH liquids the charge $U(1)$ symmetry\footnote{In fact AFQH liquids also have other spatial symmetries such as rotational symmetry on a sphere, which is characterized by a spin vector\cite{Wen1995} ${\bf s}$. In this paper we'll focus on on-site symmetries such as $U(1)$ charge conservation.} is already taken into account by coupling the ``electron" currents to an external electromagnetic gauge field $A_\mu$ with charge vector ${\bf t}$. The stability of corresponding edge states in the presence of $U(1)$ charge conservation is also studied\cite{Haldane1995}. In \Ref{Lu2012a} it was shown that under a generic unitary (antiunitary) onsite symmetry $h$ the chiral bosons $\{\phi_I\}$ in (\ref{edge lagrangian}) transform as
\begin{eqnarray}
\phi_I\rightarrow\eta_h\sum_J{\bf W}_{I,J}^h\phi_J+\delta\phi_I^h\label{symmetry:edge}
\end{eqnarray}
where $\eta_h=\pm1$ for a unitary (antiunitary) symmetry $h\in G$ ($G$ denotes the symmetry group). $\delta\phi_I^h\in[0,2\pi)$ is a $U(1)$ phase shift for the chiral bosons and ${\bf W}\in GL(N,\mathbb{Z})$ is a $N\times N$ unimodular matrix satisfying
\begin{eqnarray}
{\bf K}=\eta_h\big({\bf W}^h\big)^T{\bf K}{\bf W}^h.
\end{eqnarray}\\

For example when $G=U(1)$ \ie with conservation of $U(1)$ boson charge\cite{Levin2012a,Lu2012a}, there are integer ($\mbz$) classes of different bosonic SRE phases. They are described by the same $2\times2$ ${\bf K}$ matrix (\ref{K mat:boson}) and their edge chiral bosons have the following symmetry transformations:
\begin{eqnarray}
U_{\Delta\theta}:~~~\begin{pmatrix}\phi_1\\ \phi_2\end{pmatrix}\rightarrow\begin{pmatrix}\phi_1\\ \phi_2\end{pmatrix}+\Delta\theta\begin{pmatrix}1\\q\end{pmatrix},~~~\Delta\theta\in[0,2\pi).\label{symmetry:U(1)}
\end{eqnarray}
where $U_{\Delta\theta}$ denotes an element of Abelian symmetry group $U(1)=\{U_{\alpha}|0\leq\alpha<2\pi;U_0=U_{2\pi}=\bse;U_aU_b=U_{a+b\mod2\pi}\}$. Here $\bse$ stands for the identity element of the symmetry group. Different values of integer $q\in\mbz$ lead to distinct bosonic SRE phases which preserve $U(1)$ symmetry. From edge theory (\ref{edge lagrangian}) one can easily see that when $q=0$, the edge states can be gapped out without breaking the $U(1)$ symmetry\cite{Haldane1995}. More specifically while preserving $U(1)$ symmetry, one can add a $\cos(\phi_2)$ term to (\ref{edge lagrangian}) and pin the chiral boson field $\phi_2$ to a classical value $\langle\phi_2\rangle=const$. Hence $q=0$ corresponds to nothing but a trivial (Mott) insulator. When $q\neq0$ however, the edge state cannot be gapped out without breaking the $U(1)$ symmetry. Hence each $q\neq0$ in (\ref{symmetry:U(1)}) corresponds to a $U(1)$-SPT phase. From bulk-edge correspondence between (\ref{bulk lagrangian}) and (\ref{edge lagrangian}) it's easy to figure out\cite{Lu2012a} the global $U(1)$ symmetry (\ref{symmetry:U(1)}) corresponds to charge vector ${\bf t}=(q,1)^T$ in (\ref{bulk lagrangian}). The topological invariant characterizing these bosonic U(1) SPT phases is an \emph{even}-integer Hall conductance\cite{Lu2012a,Senthil2013,Liu2013,Chen2012}
\begin{eqnarray}
\sigma_{xy}={\bf t}^T{\bf K}^{-1}{\bf t}=2q.
\end{eqnarray}
in unit of $e_b^2/h$ ($e_b$ is the unit charge carried by bosons). Different $U(1)$-symmetric bosonic SPT phases are featured by their different Hall conductances.\\
The bulk effective theory (\ref{bulk lagrangian}) for $U(1)$-SPT phases, however, is not unique. In fact two seemingly different bulk effective theories with $U(1)$ charge conservation, labeled by $({\bf K},{\bf t})$ and $({\bf K}^\prime,{\bf t}^\prime)$, describe the same phase if they are related by the following $GL(N,\mbz)$ transformation\cite{Wen1995,Lu2012a}:
\begin{eqnarray}\label{GL(N,Z) transf}
{\bf K}^\prime={\bf X}^T{\bf K}{\bf X},~~{\bf t}^\prime={\bf X}^T{\bf t},~~~~{\bf X}\in GL(N,\mbz).
\end{eqnarray}
Therefore bosonic $U(1)$-SPT phases with ${\bf K}=\begin{pmatrix}0&1\\1&0\end{pmatrix}$ and ${\bf t}=(q,1)^T$, we can choose the following $GL(N,\mbz)$ transformation
\begin{eqnarray}
{\bf X}=\begin{pmatrix}1&0\\-q&1\end{pmatrix}\notag
\end{eqnarray}
to obtain an alternative description (\ref{bulk lagrangian}) of the same $U(1)$-SPT phase where
\begin{eqnarray}\label{K mat:boson:U(1)}
{\bf K}^\prime=\begin{pmatrix}-2q&1\\1&0\end{pmatrix},~~~{\bf t}^\prime=\begin{pmatrix}0\\1\end{pmatrix},~~~q\in\mbz.
\end{eqnarray}
In this equivalent description only the 2nd type of bosons (with currents $J^2_\mu$) carry unit $U(1)$ charge. And quantum phase transitions between two different $U(1)$-SPT phases can be conveniently discussed using this representation, similar to quantum phase transitions between different FQH liquids in the clean limit\cite{Wen1993,Read2000,Wen2000}. In the following we use (\ref{K mat:boson:U(1)}) to describe bosonic $U(1)$-SPT phases.

\section{Projective construction of bosonic SPT phases}\label{PROJECTIVE CONSTRUCTION}

A systematic way to connect a low-energy effective theory to a many-body wavefunction is through the projective (or slave particle/parton) construction\cite{ANDERSON1987,Baskaran1988,Affleck1988a,Kotliar1988,Read1991,Sachdev1992,Mudry1994,Wen1996,Wen1999,Wen2002}. In a projective construction, usually the physical microscopic degrees of freedom (\eg spins or electrons) are written in terms of ``fractionalized" degree of freedom (\dof) called ``partons". The Hilbert space of partons are larger than the original physical Hilbert space, and a projection onto the physical Hilbert space is needed to get rid of spurious \dof. As a result the low-energy theory involves partons couple to gauge fields. The gauge fields serve as the glue which bind the partons together to form a physical microscopic \dof (or to enforce the local constraints). First one construct the mean-field state of partons in the enlarged Hilbert space which determines the gauge structure\cite{Wen1999,Wen2002}. The physical many-body wavefunction is then obtained by projecting the parton mean-field state onto the physical Hilbert space. Projective construction is a powerful tool which allows one to write down many-body wavefunctions for a quantum phase, once we know its long-wavelength effective theory.

In a projective construction each parton usually carries a fraction of the physical quantum numbers, such as fractional charge\cite{Laughlin1983} or fractional statistics\cite{Arovas1984} in $2+1$-D. Since there is no fractional quasiparticles in SPT phases\cite{Chen2013}, at the first sight it seems that projective construction  will not be so uesful. However this turns out to be not true. In the following we demonstrate the value of projective construction for describing bosonic $U(1)$-SPT. This formalism can be easily generalized to bosonic/fermionic SPT phases in two space dimensions with other symmetries.


\subsection{The projective construction and effective theory}

The many-body wavefunction\cite{Laughlin1983,Halperin1983} for a AFQH liquid described by effective theory (\ref{bulk lagrangian}) is
\begin{eqnarray}\label{wf:AFQH}
\Psi_{\bf K}=\prod_{i<j,I,J}\big(z_i^{(I)}-z_j^{(J)}\big)^{{\bf K}_{I,J}}e^{-\sum_{i,I}|z_i^{(I)}|^2/4}.
\end{eqnarray}
if all electrons stay in the lowest Landau level (LLL) on a disc. $z^I_j=x^I_j+\imth y^I_j$ denotes the two-dimensional complex coordinate of the $j$-th electron of the $I$-th type. It's easy to check the anyonic excitations\cite{Laughlin1983,Arovas1984,Halperin1984,Read1990} on top of such a ground state wavefunction (\ref{wf:AFQH}) are fully captured by Chern-Simons theory (\ref{bulk lagrangian}). Notice that the wavefunction for a filled LLL of spinless fermions $\{z_i=x_i+\imth y_i\}$ with charge $Q$ (in unit of electron charge $e$) is the following Vandermonde determinant
\begin{eqnarray}
&\notag\Psi_{LLL}\{z_i\}=\det\Big\{{\bf M}_{i,j}=z_i^{j-1}e^{-Q|z_i|^2/4}\Big\}\\
&\label{wf:LLL}=\prod_{i<j}(z_i-z_j)^1\cdot e^{-Q\sum_i|z_i|^2/4}
\end{eqnarray}
The idea of projective construction is to split each electron into several partons\cite{Jain1989a,Wen1991b} (each carrying fractional charge $Q_\alpha$ with $\sum_\alpha Q_\alpha=1$), and each type of partons fill the LLL\footnote{In more general cases certain types of partons could form integer quantum Hall liquids by filling several Landau levels.} for the simplest case. Let's consider a Laughlin state as an illustration\cite{Wen1998}:
\begin{eqnarray}
&\notag\Psi_{Laughlin}=\prod_{i<j}(z_i-z_j)^m\cdot e^{-\sum_i|z_i|^2/4}\\
&=\langle0|\prod_{i}\big[\prod_{\alpha=1}^mf_\alpha(z_i)\big]|MF\rangle.\notag
\end{eqnarray}
where $|MF\rangle$ is the mean-field state where each type of fermionic partons $f_\alpha$ fill a LLL. $|0\rangle$ is the parton vacuum and each parton carries charge $Q_\alpha=\frac1m$. And the electron operator $c$ is the product of all parton operators $f_\alpha$ as
\begin{eqnarray}
c(z)=\prod_{\alpha=1}^mf_\alpha(z).\notag
\end{eqnarray}\\

The above projective construction can be generalized to an arbitrary ${\bf K}$ matrix with effective theory (\ref{bulk lagrangian}) and wavefunction (\ref{wf:AFQH}). The many-body wavefunction can be obtained by the following projection of the parton mean-field state $|MF\rangle$:
\begin{eqnarray}\label{wf:proj}
\Psi_{\bf K}\Big(\{{\bf r}_j^I\}\Big)=\langle0|\prod_I\prod_{j}c_I({\bf r}^I_j)|MF\rangle.
\end{eqnarray}
where $c_I({\bf r})$ is the annihilation operator for electrons of the $I$-th type at coordinate ${\bf r}=(x,y)$, written in terms of fermionic parton operators $\{f_\alpha\}$. ${\bf r}_j^I$ is the position of $j$-th electron of the $I$-th type. Restricting ourselves to a generic $2\times2$ matrix ${\bf K}$:
\begin{eqnarray}\label{K mat:2x2}
{\bf K}=\begin{pmatrix}n+l&n\\n&n+m\end{pmatrix},~~~l,m,n\in\mbz.
\end{eqnarray}
the two types of electron operators can be written as
\begin{eqnarray}
&\notag c_1(z)=\prod_{\alpha=1}^{|n|}f_\alpha(z)\cdot\prod_{\beta=1}^{|l|}f_{\uparrow,\beta}(z),\\
&c_2(z)=\prod_{\alpha=1}^{|n|}f_\alpha(z)\cdot\prod_{\gamma=1}^{|m|}f_{\downarrow,\gamma}(z).\label{parton:2x2}
\end{eqnarray}
where $\{f_\alpha|1\leq\alpha\leq|n|\}$, $\{f_{\uparrow,\beta}|1\leq\beta\leq|l|\}$ and $\{f_{\downarrow,\gamma}|1\leq\gamma\leq|m|\}$ are the $|l|+|m|+|n|$ different types of fermionic partons. If $n>0$, then each $f_\alpha$-parton fills a LLL with wavefunction (\ref{wf:LLL}) in the mean-field state. If $n<0$ on the other hand, each $f_\alpha$-parton fills a LLL under the opposite magnetic field with wavefunction
\begin{eqnarray}
&\Psi_{\bar{LLL}}\{z_i\}\label{wf:bar-LLL}=\prod_{i<j}(\bar{z}_i-\bar{z}_j)^1\cdot e^{-Q\sum_i|z_i|^2/4}
\end{eqnarray}
Here $\bar{z}=x-\imth y$ represents complex conjugation of $z$. Similarly each type of $f_{\uparrow,\beta}$, $f_{\downarrow,\gamma}$ partons fills a LLL, under a magnetic field depending on the sign of $l$ and $m$. Here $\uparrow,\downarrow$ have nothing to do with electron spins.

The local constraints which enforce the projection into physical Hilbert space is the following:
\begin{eqnarray}
&\notag f^\dagger_{1}(z)f_{1}(z)=f^\dagger_{2}(z)f_2(z)=\cdots=f^\dagger_{|n|}(z)f_{|n|}(z),\\
&\notag f^\dagger_{\uparrow,1}(z)f_{\uparrow,1}(z)=\cdots=f^\dagger_{\uparrow,|l|}(z)f_{\uparrow,|l|}(z),\\
&\notag f^\dagger_{\downarrow,1}(z)f_{\downarrow,1}(z)=\cdots=f^\dagger_{\downarrow,|m|}(z)f_{\downarrow,|m|}(z),\\
&f^\dagger_{1}(z)f_{1}(z)=f^\dagger_{\uparrow,1}(z)f_{\uparrow,1}(z)+f^\dagger_{\downarrow,1}(z)f_{\downarrow,1}(z).\label{constraint:2x2}
\end{eqnarray}
This is straightforward to see from the parton construction (\ref{parton:2x2}), since the physical excitations in the system are always generated by electron operators $\{c_{1,2},c^\dagger_{1,2}\}$. Apparently when both $n+l$ and $n+m$ are even integers, $c_1$ and $c_2$ are both bosonic operators. On the other hand if one of $n+l$ and $n+m$ is an odd integer, there will be fermionic particles in the system, as discussed earlier for effective theory (\ref{bulk lagrangian}).\\

In the following we show that enforcing constraints (\ref{constraint:2x2}) in the aforementioned parton mean-field state indeed yields the effective Chern-Simons theory (\ref{bulk lagrangian})\cite{Barkeshli2012}. First let's write the dual form (\ref{duality:electron current}) of parton currents in $2+1$-D:
\begin{eqnarray}
&\notag\frac{\epsilon^{\mu\nu\lambda}}{2\pi}\partial_\nu a^\alpha_\lambda=J_\alpha^\mu\equiv\text{currents of $f_\alpha$ partons},\\
&\notag\frac{\epsilon^{\mu\nu\lambda}}{2\pi}\partial_\nu a^{\uparrow,\beta}_\lambda=J_{\uparrow,\beta}^\mu\equiv\text{currents of $f_{\uparrow,\beta}$ partons},\\
&\notag\frac{\epsilon^{\mu\nu\lambda}}{2\pi}\partial_\nu a^{\downarrow,\gamma}_\lambda=J_{\downarrow,\gamma}^\mu\equiv\text{currents of $f_{\downarrow,\gamma}$ partons}.
\end{eqnarray}
In terms of $U(1)$ gauge fields $\{a_\mu^\alpha,a_\mu^{\uparrow,\beta},a_\mu^{\downarrow,\gamma}\}$ the effective theory corresponding to the parton mean-field states of filled LLLs is\cite{Wen1995}
\begin{eqnarray}
&\notag\mathcal{L}_{MF}=-\frac{\text{Sgn}(n)}{4\pi}\epsilon^{\mu\nu\lambda}\sum_{\alpha=1}^{|n|}a_\mu^\alpha\partial_\nu a^\alpha_\lambda-\\
&\frac{\text{Sgn}(l)}{4\pi}\epsilon^{\mu\nu\lambda}\sum_{\beta=1}^{|l|}a_\mu^{\uparrow,\beta}\partial_\nu a^{\uparrow,\beta}_\lambda-\frac{\text{Sgn}(m)}{4\pi}\epsilon^{\mu\nu\lambda}\sum_{\gamma=1}^{|m|}a_\mu^{\downarrow,\gamma}\partial_\nu a^{\downarrow,\gamma}_\lambda.\notag
\end{eqnarray}
This is the effective theory description of filled Landau levels. The local constraints (\ref{constraint:2x2}) now can be enforced by introducing Lagrangian multipliers $\{b_\mu,b_\mu^\alpha,b_\mu^{\uparrow,\beta},b_\mu^{\downarrow,\gamma}\}$ in a covariant way\cite{Barkeshli2012}:
\begin{eqnarray}
&\mathcal{L}_{eff}=\mathcal{L}_{MF}+\mathcal{L}_{constraint},\notag\\
&\notag\mathcal{L}_{constraint}=\frac{\epsilon^{\mu\nu\lambda}}{4\pi}\sum_{\alpha=1}^{|n|-1}b_\mu^\alpha\partial_\nu(a^\alpha_\lambda-a^{\alpha+1}_\lambda)+\\
&\notag\frac{\epsilon^{\mu\nu\lambda}}{4\pi}\sum_{\beta=1}^{|l|-1}b_\mu^{\uparrow,\beta}\partial_\nu (a^{\uparrow,\beta}_\lambda-a^{\uparrow,\beta+1}_\lambda)\\
&\notag+\frac{\epsilon^{\mu\nu\lambda}}{4\pi}\sum_{\gamma=1}^{|m|-1}b_\mu^{\downarrow,\gamma}\partial_\nu (a^{\downarrow,\gamma}_\lambda-a^{\downarrow,\gamma+1}_\lambda)\\
&\notag+\frac{\epsilon^{\mu\nu\lambda}}{4\pi}b_\mu\partial_\nu(a_\lambda^1-a_\lambda^{\uparrow,1}-a_\lambda^{\downarrow,1}).
\end{eqnarray}
After integrating out the gauge fields $\{b_\mu,b_\mu^\alpha,b_\mu^{\uparrow,\beta},b_\mu^{\downarrow,\gamma}\}$ the constraints\footnote{This is just a covariant form of constraints (\ref{constraint:2x2}) shown earlier.} is enforced:
\begin{eqnarray}
&\notag J_1=\cdots=J_{|n|},~~~J_1=J_{\uparrow,1}+J_{\downarrow,1},\\
&\notag J_{\uparrow,1}=\cdots=J_{\uparrow,|l|},~~~J_{\downarrow,1}=\cdots=J_{\downarrow,|m|}.
\end{eqnarray}
Solving these constraints we obtain the Chern-Simons effective theory (\ref{bulk lagrangian}) with $2\times2$ ${\bf K}$ matrix (\ref{K mat:2x2}):
\begin{eqnarray}
&\notag\mathcal{L}_{eff}=-\frac{n\epsilon^{\mu\nu\lambda}}{4\pi}\big(a_\mu^{\uparrow,1}\partial_\nu a^{\downarrow,1}_\lambda+a_\mu^{\downarrow,1}\partial_\nu a^{\uparrow,1}_\lambda\big)-\\
&\frac{(n+l)\epsilon^{\mu\nu\lambda}}{4\pi}a_\mu^{\uparrow,1}\partial_\nu a^{\uparrow,1}_\lambda-\frac{(n+m)\epsilon^{\mu\nu\lambda}}{4\pi}a_\mu^{\downarrow,1}\partial_\nu a^{\downarrow,1}_\lambda.\notag
\end{eqnarray}

\subsection{Constructing many-body wavefunctions of bosonic SPT phases on a lattice}

Based on general discussions in previous sections, now we focus on bosonic $U(1)$-symmetric SPT phases in $2+1$-D. They are described by Chern-Simons effective theory (\ref{bulk lagrangian}) with $2\times2$ matrix ${\bf K}=\begin{pmatrix}-2q&1\\1&0\end{pmatrix}$ and charge vector ${\bf t}=(0,1)^T$ (see (\ref{K mat:boson:U(1)})). Following procedures discussed earlier, we can systematically write down their many-body wavefunctions. There are two types of bosons in the system, whose annihilation operators at position ${\bf r}=(x,y)$ are
\begin{eqnarray}
&\notag b_1({\bf r})=f({\bf r})\prod_{\alpha=1}^{|2q+1|}d_\alpha({\bf r}),\\
&b_2({\bf r})=f({\bf r})f_1({\bf r})\label{parton:u(1) spt}.
\end{eqnarray}
where $f,f_{1}$ and $\{d_\alpha,~1\leq\alpha\leq|2q+1|\}$ are all fermionic partons. Here ``electron'' operators $b_{1,2}$ correspond to $c_{1,2}$ in the general case (\ref{parton:2x2}), while the partons used in (\ref{parton:u(1) spt}) and previously (\ref{parton:2x2}) have the following correspondence:
\bea
&\notag f\longrightarrow f_\alpha,\\
&\notag d_\alpha\longrightarrow f_{\uparrow,\beta},\\
&\notag f_1\longrightarrow f_{\downarrow,\gamma}.
\eea
Apparently $b_{1,2}$ are both bosonic operators since they contain an even number of fermionic operators. In the mean-field state we require $f$-partons fill a LLL with wavefunction (\ref{wf:LLL}) and $f_1$-partons fill a LLL under an opposite magnetic field with wavefunction (\ref{wf:bar-LLL}). Meanwhile each type of $d_\alpha$-partons will also fill a LLL, under a magnetic field whose direction depends on the sign of $2q+1$. If $2q+1>0$ then each type of $d_\alpha$-partons fill a LLL with wavefunction (\ref{wf:bar-LLL}); if $2q+1<0$ each type fill a LLL with wavefunction (\ref{wf:LLL}). Even if all partons are fermionic in this projective construction (\ref{parton:u(1) spt}), all physical excitations on top of many-body ground state (\ref{wf:proj}) are always bosonic. This is because whenever a fermionic parton is excited, it is always combined with certain $U(1)$ gauge fluxes which enforce the constraints (\ref{constraint:2x2}) in the physical spectrum. Therefore although we build physical bosons out of fermionic partons, there is no true fermion excitations or any fractionalization.

So far we've been discussing many-body wavefunctions of variables $(x,y)$ in continuous two-dimensional space. Their building block is the filled-LLL wavefunction (\ref{wf:LLL}) or (\ref{wf:bar-LLL}). A natural question is: how to construct a similar many-body wavefunction whose variables live on a lattice? It turns out the projective construction can be applied straightforwardly to the lattice case\cite{Lu2012,McGreevy2012}: the key change is to replace the filled LLL by a filled band\cite{Hofstadter1976,Haldane1988} with Chern number\cite{Thouless1982} $\pm1$. Then projective construction leads to the same low-energy effective theory (\ref{bulk lagrangian}) in the long wavelength limit.

The major complications for implementing projective construction on a lattice are two-fold. Firstly, because of the local constraints (\ref{parton:u(1) spt}):
\begin{eqnarray}
&\notag d^\dagger_1({\bf r})d_1({\bf r})=\cdots=d^\dagger_{|2q+1|}({\bf r})d_{|2q+1|}({\bf r}),\\
&f^\dagger({\bf r})f({\bf r})=f_1^\dagger({\bf r})f_1({\bf r})+d_1^\dagger({\bf r})d_1({\bf r}).\label{constraint:u(1) spt}
\end{eqnarray}
where ${\bf r}$ denotes an orbital on a lattice site. Hence the filling fractions (per unit cell, say) of fermionic partons satisfy
\begin{eqnarray}
&\notag\nu_{d_1}=\nu_{d_2}=\cdots=\nu_{d_{|2q+1|}}=\nu_{b_1},\\
&\label{constraint:filling}\nu_{f}=\nu_{f_1}+\nu_{d_1}=\nu_{b_1}+\nu_{b_2}.
\end{eqnarray}
Notice that for a many-body wavefunction described by effective theory (\ref{bulk lagrangian}), its corresponding mean-field state requires each type of partons to form a band insulator with desired Chern number $\pm1$. One has to choose the hopping parameters so that the resultant state is an insulator with correct fillings (\ref{constraint:filling}) and Chern numbers
\begin{eqnarray}
&\notag C_{d_1}=C_{d_2}=\cdots=C_{d_{|2q+1|}}=-\text{Sgn}(2q+1),\\
& C_{f_1}=-1,~~~C_{f}=+1.\label{constraint:chern number}
\end{eqnarray}
%

Secondly, a lattice structure comes in together with certain space group symmetries. In the presence of lattice symmetry, effective theory (\ref{bulk lagrangian}) alone is not enough to fully characterize all different topological phases from the projective construction\cite{Wen2002,Lu2012}. Specifically there can be many different phases with the same effective theory (\ref{bulk lagrangian}), which carry different quantum numbers of lattice symmetry. Different universality classes of parton mean-field ansatz\cite{Wen2002,Lu2012} (but with the same effective theory), correspond to these different topological phases distinguished by lattice symmetry. In this work however, we will not attempt to classify all the different $U(1)$-SPT phases with certain lattice symmetries. Instead we'll show a simple parton mean-field ansatz on square lattice, which gives rise to many-body wavefunctions of bosonic $U(1)$-SPT phases by projective construction (\ref{wf:proj}).

\subsection{An example of bosonic $U(1)$-SPT phases on square lattice at half-filling: $\nu_{b_1}=\nu_{b_2}=1/2$}

We use a simple example to illustrate the projective construction of bosonic SPT phases on a lattice. We consider bosonic $U(1)$-SPT phases, which is described by effective theory (\ref{bulk lagrangian}) with ${\bf K}$ matrix (\ref{K mat:boson:U(1)}). In this example there are two types of bosons $\{b_{1},b_2\}$ \big(see (\ref{parton:u(1) spt})\big) living on a square lattice. We choose each type of bosons to have filling fraction $\nu_{b_1}=\nu_{b_2}=\frac12$, \ie on average there are one boson (of each type) per two sites. Since the charge vector is ${\bf t}=(0,1)^T$, each boson of the 2nd type carries a unit of charge while bosons of 1st type carry no charges. One can think that the two types of bosons stay in two different orbitals (say, labeled by pseudospin $\uparrow/\downarrow$) respectively on every lattice site. In the corresponding projective construction (\ref{parton:u(1) spt}), fermionic partons $\{d_\alpha,~1\leq\alpha\leq|2q+1|\}$ and $f_1$ all have the same filling fraction $\nu=\frac12$ just as bosons $b_1$ and $b_2$. On the other hand, as shown in (\ref{constraint:filling}) the filling fraction for fermionic $f$-partons is $\nu_f=\nu_{b_1}+\nu_{b_2}=1$, \ie on average there is one $f$-parton per site on the square lattice. All $d_\alpha$-partons stay on $\uparrow$-orbitals, while all $f_1$-partons stay on $\downarrow$-orbitals. Meanwhile $f$-partons can hop on both orbits in every site, whose filling fraction is twice as much as $f_1$- and $d_\alpha$-partons.

\begin{figure}
 \includegraphics[width=0.4\textwidth]{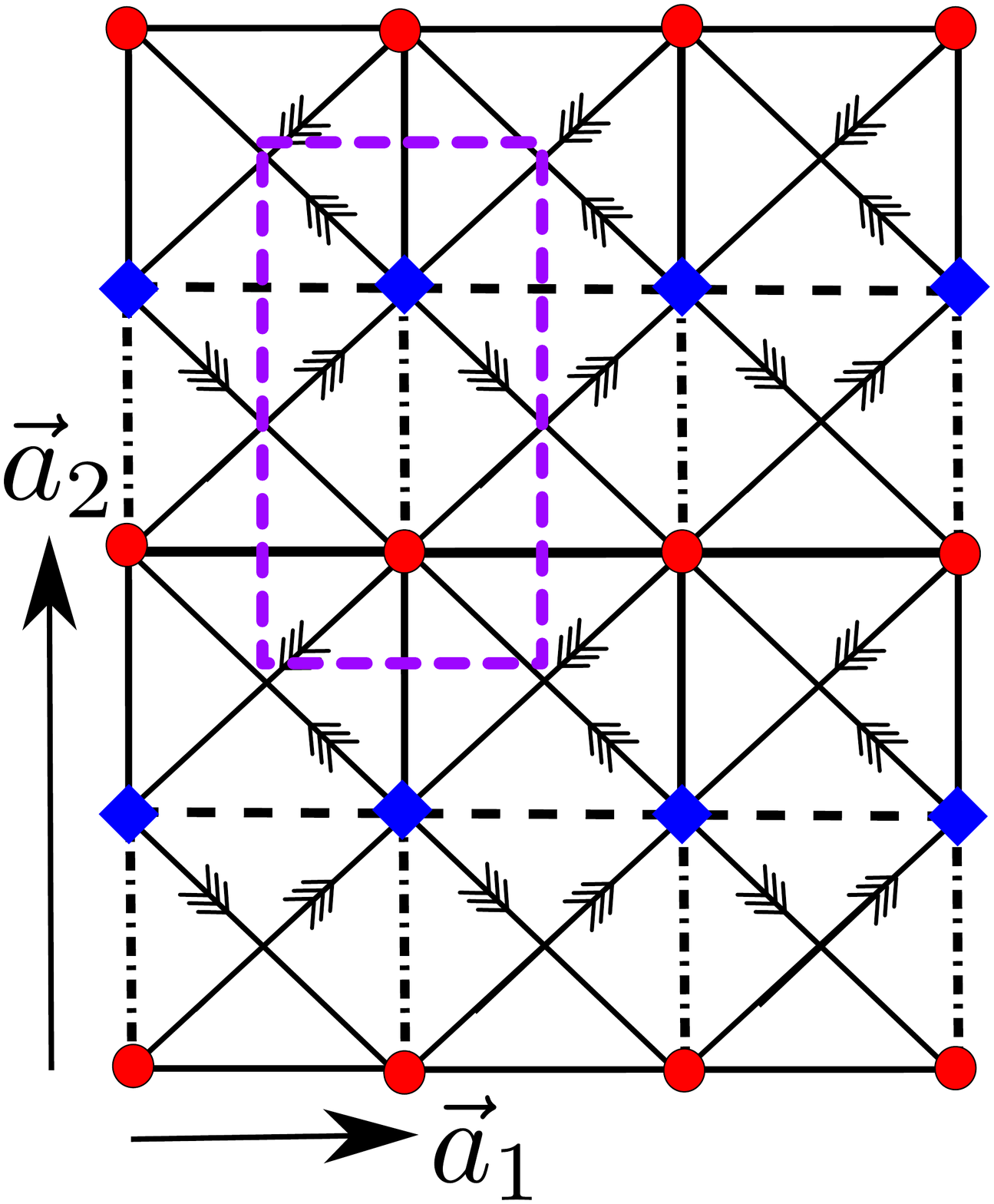}
\caption{(color online) An illustration of mean-field hopping ansatz of $f$- and $d_\alpha$-partons on square lattice. The magnetic unit cell of the mean-field Hamiltonian contains two lattice sites (featured by red circles and blue diamonds respectively) of the original square lattice, as indicated by the pink rectangle. The primitive vectors $\vec{a}_{1,2}$ for the magnetic unit cell are also shown. For simplicity only 1st and 2nd nearest neighbor (NN) hoppings are shown here. Among \emph{horizontal} hoppings between 1st NNs, solid lines denote hopping parameter $t_x$ while dashed lines denote $-t_x$ where $t_x>0$ is a real number. For \emph{vertical} hopping between 1st NNs, solid lines denote hopping parameter $t_y$ while dashed lines denote $t_y^\prime$ (we choose $t_y,t_y^\prime>0$ for simplicity). The 2nd neighbor hopping parameters are all imaginary and equals $\imth t_2$ along the arrow directions ($t_2>0$). This hopping Hamiltonian in momentum space is given by (\ref{mf hopping:f1,d}).}\label{fig:pi-flux hopping}
\end{figure}

Since $f_1$- and $d_\alpha$-partons all have a filling fraction $\nu=\frac12$, we need to enlarge the unit cell in their mean-field hopping Hamiltonians to guarantee that every type of partons can form a band insulator. In the case of half-filling ($\nu=\frac12$), a $\pi$-flux needs to be inserted into each plaquette on square lattice to \emph{double the unit cell} in the mean-field Hamiltonian, and in the presence of a gap each type of $f_1$-, $d_\alpha$-partons fill the lower band. To construct a bosonic $U(1)$-SPT phase with (\ref{K mat:boson:U(1)}), we need their Chern numbers to satisfy condition (\ref{constraint:chern number}). This can be realized by the hopping Hamiltonian of spinless fermions depicted in FIG. \ref{fig:pi-flux hopping}.

With the doubled unit cell for the mean-field hopping ansatz shown in FIG. \ref{fig:pi-flux hopping}, one can label momentum ${\bf k}=k_1\vec{b}_1+k_2\vec{b}_2$ where $\vec{b}_{1,2}$ are primitive reciprocal lattice vectors associated with Bravais lattice vectors $\vec{a}_{1,2}$ in FIG. \ref{fig:pi-flux hopping}. After a Fourier transformation the mean-field hopping Hamiltonian is a functional of parameters $(t_x,t_y,t_y^\prime,t_2)$:
\begin{eqnarray}
&\notag\mathcal{H}_{\bf k}\big(t_x,t_y,t_y^\prime,t_2\Big)=\begin{bmatrix}2t_x\cos{k_1}&t_y^\prime+t_y e^{-\imth k_2}\\t_y^\prime+t_y e^{\imth k_2}&-2t_x\cos{k_1}\end{bmatrix}\\
&+2t_2\sin{k_1}\begin{bmatrix}0&1-e^{-\imth k_2}\\1-e^{\imth k_2}&0\end{bmatrix}.\label{mf hopping:f1,d}
\end{eqnarray}
where the column and row indices label the two sites in each unit cell (red circles and blue diamonds in FIG. \ref{fig:pi-flux hopping}) of FIG. \ref{fig:pi-flux hopping}. The 1st line of (\ref{mf hopping:f1,d}) corresponds to real hoppings between 1st NNs while the 2nd line is associated with imaginary hoppings between 2nd NNs. In the special case $t_y=t_y^\prime$ and $t_2=0$, the band structure (\ref{mf hopping:f1,d}) has two Dirac nodes at $(k_1,k_2)=(\pm\pi/2,\pi)$:
\begin{eqnarray}
\mathcal{H}_{\bf k}\approx\mp2t_x\sigma_z\delta k_1-t_y\sigma_y\delta k_2,~~~\delta{\bf k}={\bf k}\mp(\frac{\pi}{2},\pi).\label{mf hopping:Dirac fermions}
\end{eqnarray}
where $\sigma_{x,y,z}$ denote the three Pauli matrices. These Dirac cones is a generic feature of 1st-NN-hopping-only $\pi$-flux states\cite{Affleck1988a,Marston1989} with magnetic translation symmetry\cite{Wen2002}. Now we can turn on perturbations to gap out the Dirac nodes by allowing $t_y\neq t_y^\prime$ and $t_2\neq0$. The lowest-order effect of nonzero $t_y^\prime-t_y$ and $t_2$ on Dirac fermions (\ref{mf hopping:Dirac fermions}) is to add a mass term:
\begin{eqnarray}
\mathcal{M}_{{\bf k}=(\pm\pi/2,\pi)}=(t_y^\prime-t_y\pm4t_2)\sigma_x.\notag
\end{eqnarray}
It then follows that the Chern number $C_{\mathcal{H}_{\bf k}}$ of the lower band in (\ref{mf hopping:f1,d}) is
\begin{eqnarray}
C_{\mathcal{H}_{\bf k}}=\Big\{\left.\begin{aligned}0,~~~~~~&|t_y^\prime-t_y|>4|t_2|\\
-\text{Sgn}(t_2),~~~~~~&|t_y^\prime-t_y|<4|t_2|\end{aligned}\right.\label{chern number}
\end{eqnarray}
And the band gap of mean-field hopping Hamiltonian (\ref{mf hopping:f1,d}) is of the order $\Delta_{MF}\sim\min\big\{|t_y^\prime-t_y+4t_2|,|t_y^\prime-t_y-4t_2|\big\}$.

To realize the Chern numbers (\ref{constraint:chern number}) for filled parton bands, one can simply choose the following mean-field hopping Hamiltonians for $d_\alpha$-, $f_1$-partons:
\begin{eqnarray}
&\notag H_{\bf k}^{d_\alpha}=\mathcal{H}_{\bf k}\Big(t_x^\alpha,t_x^\alpha,t_x^\alpha,|t_2|\cdot\text{Sgn}(2q+1)\Big),\\
&\notag1\leq\alpha\leq|2q+1|;\\
&\notag H_{\bf k}^{f_1}=\mathcal{H}_{\bf k}\Big(t_x,t_x,t_x,|t_2|\Big).
\end{eqnarray}
Here we can choose different 1st NN hopping parameters $t_x^\alpha$ for different $d_\alpha$-partons\footnote{This choice can break the original $SU(|2q+1|)\times U(1)$ gauge structure of projective construction (\ref{parton:u(1) spt}) down to $U(1)^{|2q+1|}$. However this detail is not important for topological properties\cite{Barkeshli2012} of many-body wavefunction (\ref{wf:proj}) obtained from projective construction (\ref{parton:u(1) spt}).}. At the mean-field level both $d_\alpha$- and $f_1$-partons fill the lower band of their hopping Hamiltonian. Note that $d_\alpha$-partons only hop on $\uparrow$-orbitals and $f_1$-partons on $\downarrow$-orbitals. On the other hand, $f$-partons hop on both orbitals, and its hopping Hamiltonian $H_{\bf k}^f$ is the following $4\times4$ matrix in momentum space:
\begin{eqnarray}
H_{\bf k}^f=\begin{bmatrix}\langle\uparrow|\hat{H}_{\bf k}^f|\uparrow\rangle&\langle\uparrow|\hat{H}_{\bf k}^f|\downarrow\rangle\\ \langle\uparrow|\hat{H}_{\bf k}^f|\downarrow\rangle^\dagger&\langle\downarrow|\hat{H}_{\bf k}^f|\downarrow\rangle\end{bmatrix}\notag
\end{eqnarray}
The $f$-partons should fill the two lower bands, with filling fraction $\nu_f=1$ per site. Therefore we choose the following hopping parameters to satisfy total Chern number $C_f=1$ for the two filled $f$-parton bands:
\begin{eqnarray}
&\notag \langle\uparrow|\hat{H}_{\bf k}^{f}|\uparrow\rangle=\mathcal{H}_{\bf k}\Big(t_x,t_x,t_x,-|t_2|\Big),\\
&\notag \langle\downarrow|\hat{H}_{\bf k}^{f}|\downarrow\rangle=\mathcal{H}_{\bf k}\Big(t_x,t_y,t_y+|t_2|,0\Big),\\
&\notag \langle\uparrow|\hat{H}_{\bf k}^{f}|\downarrow\rangle=t_0\cdot I_{2\times2}.
\end{eqnarray}
We choose a simplest form of uniform on-site hopping between $\uparrow$- and $\downarrow$-orbitals, with amplitude $t_0$. As long as $|t_0|\ll|t_2|$ it's straightforward to show that total Chern number of two filled $f$-parton bands is $C_f=1+0=1$.

The above mean-field ansatz $|MF\rangle$ for $d_\alpha$- , $f_1$- and $f$-partons yields a many-body wavefunction of bosonic $U(1)$-SPT phase with Hall conductance $\sigma_{xy}=2q$, after projection (\ref{wf:proj}) is applied. Notice that the position ${\bf r}$ in projection (\ref{wf:proj}) in this case corresponds to the lattice site position ${\bf r}=x\vec{a}_1+y\vec{a}_2/2\equiv(x,y)$. The many-body wavefunction with two-types of bosons $\{b_1({\bf r}^1_i)\}$ and $\{b_2({\bf r}^2_j)\}$ is given by
\begin{eqnarray}
&\label{wf:u(1) spt}\Psi_{SPT}\Big(\{{\bf r}^1_i\};\{{\bf r}^2_j\}\Big)=\\
&\notag\langle0|\prod_{i,j}\Big(\prod_{\alpha=1}^{|2q+1|}d_\alpha({\bf r}_i^1)\Big)f_1({\bf r}_j^2)f({\bf r}_i^1,\uparrow)f({\bf r}_j^2,\downarrow)|MF\rangle.
\end{eqnarray}

Can we find out a microscopic model of interacting bosons, which realizes the above many-body wavefunction (\ref{wf:u(1) spt}) as its ground state? The answer is yes. In fact there is a systematic way to reverse engineer the interacting boson model from the parton band structure\cite{Motrunich2007,McGreevy2012}. The strategy is to enforce the ``hard'' on-site constraint (\ref{constraint:u(1) spt}) by ``soft'' energy penalty such as
\bea
\hat{H}_U=U\sum_{\bf r}\big[f^\dagger({\bf r})f({\bf r})-f_1^\dagger({\bf r})f_1({\bf r})-d_1^\dagger({\bf r})d_1({\bf r})\big]^2\notag
\eea
\eg in the case of $q=0$. Since the physical Hilbert space for hard-core bosons $b_{1,2}$ is nothing but the ground states manifold of $\hat{H}_U$, we can obtain a boson Hamiltonian by degenerate perturbation theory in $t/U$ expansion\cite{Lu2014}, where parton hopping terms (\ref{mf hopping:f1,d}) serve as perturbations to $\hat{H}_U$. In particular, the leading order terms\cite{Lu2014} are boson hoppings between 1st and 2nd NNs, as well as 1st/2nd NN repulsive interactions between bosons. For example, the hopping amplitudes of boson $b_1$ are proportional to the product of parton hopping amplitudes of $f$ and $d_1$.

\section{Quantum phase transitions between bosonic SPT phases}\label{QUANTUM PHASE TRANSITION}

The projective construction (\ref{parton:u(1) spt}) not only provides the variational many-body wavefunction (\ref{wf:u(1) spt}), but also allows us to study the phase transitions between different SPT phases in two dimensions. Here we'll again focus on $U(1)$-symmetric SPT phases of bosons $b_{1,2}$, featured by Hall conductance $\sigma_{xy}=2q$.

\subsection{Continuous quantum phase transitions between two different bosonic $U(1)$-SPT phases: emergent $QED_3$ with $N_f=2$}

In deriving of effective theory (\ref{bulk lagrangian}) from projective construction (\ref{parton:u(1) spt}), it's straightforward to see that whenever the total Chern number $C_d\equiv\sum_{\alpha=1}^{|2q+1|} C_{d_\alpha}$ of partons $\{d_\alpha,~1\leq\alpha\leq|2q+1|\}$ changes by $2p$ in the parton mean-field state $|MF\rangle$, the Hall conductance $\sigma_{xy}$ would accordingly change by $2p$. This is because the ${\bf K}$ matrix in effective theory (\ref{bulk lagrangian}) is changed from ${\bf K}=\begin{pmatrix}-2q&1\\1&0\end{pmatrix}$ to ${\bf K}^\prime=\begin{pmatrix}2(-q\pm p)&1\\1&0\end{pmatrix}$. Meanwhile the charge vector ${\bf t}=(0,1)^T$ in (\ref{bulk lagrangian}) remains the same. Therefore the quantum phase transitions between different $U(1)$-SPT phases are realized by the change of Chern number
\begin{eqnarray}
C_d\equiv\sum_{\alpha=1}^{|2q+1|}C_{d_\alpha}=~\text{odd integer}.
\end{eqnarray}
by an even integer in the $d_\alpha$-parton mean-field state. Such a Chern number changing process can happen when certain fermion mass terms change sign at several quadratic band touching (QBT) points or Dirac points. At the critical point these parton bilinear mass terms vanish, and the system is described by gapless fermions (either at QBT points or Dirac points) coupled with emergent gauge fields. However a QBT point is known\cite{Sun2009} to be marginally unstable against four-fermion repulsive interaction. Therefore a stable critical point between two different $U(1)$-SPT phases are described by mass changing sign at an \emph{even} number of Dirac nodes in the presence of dynamical gauge fields.

Let's first look at a simplest example where $q=0$ in projective construction (\ref{parton:u(1) spt}). Still we keep $C_{f_1}=-1$ and $C_f=+1$ in the parton mean-field state. If the Chern number of the filled $d_1$-parton band is $C_{d_1}=\pm1$, the corresponding effective theory (\ref{bulk lagrangian}) has ${\bf K}=\begin{pmatrix}1\pm1&1\\1&0\end{pmatrix}$. To be specific, in the mean-field Hamiltonian $H_{\bf k}^{d_1}=\mathcal{H}_{\bf k}\big(t_x,t_x,t_x,t_2\big)$ for $d_1$-parton, when the 2nd NN hopping parameter $t_2$ in FIG. \ref{fig:pi-flux hopping} changes sign from positive to negative, the Chern number of the filled lower band of $d_1$-partons will change from $-1$ to $+1$. And this realizes a continuous quantum phase transition between the trivial boson insulator ($\sigma_{xy}=0$) and a bosonic $U(1)$-SPT phase with $\sigma_{xy}=-2$ (in unit of $1/2\pi$). The effective theory describing the above continuous phase transition is
\begin{eqnarray}
&\notag\mathcal{L}_{qpt}=\sum_{s=1}^2\bar{\psi}_s\gamma^\mu(\imth\partial_\mu+b_\mu)\psi_s+m\sum_{s=1}^2\bar{\psi}_s\psi_s\\
&\frac{\epsilon^{\mu\nu\lambda}}{4\pi}\Big[2b_\mu\partial_\nu(a^f_\lambda-a^{f_1}_\lambda)+a^{f_1}_\mu\partial_\nu a^{f_1}_\lambda-a^{f}_\mu\partial_\nu a^{f}_\lambda\Big]\notag\\
&-\frac1{2\pi}\epsilon^{\mu\nu\lambda}A_\mu\partial_\nu a^{f_1}_\lambda.\label{lagrangian:nf=2}
\end{eqnarray}
as indicated by the Dirac spectrum\footnote{The fermi velocity anisotropy ($v_x=2|t_x|$ and $v_y=|t_y|$) is ignored in the effective theory here. \Ref{Hermele2005} shows that the velocity anisotropy is actually \emph{irrelevant} for (\ref{QED3:Nf=2}).} (\ref{mf hopping:Dirac fermions}) of $H_{\bf k}^{d_1}$ when $t_2=0$. Here $\bar{\psi}_s=\psi_s^\dagger\sigma_x$ and $\gamma^0=\sigma_x,\gamma^x=-\sigma_y,\gamma^y=\sigma_z$. Two-component Dirac spinor $\psi_s$ with $s=1,2$ are low-energy $d_1$-parton modes around the two Dirac cones at ${\bf k}=(\pm\pi/2,\pi)$ in band structure (\ref{mf hopping:Dirac fermions}). Notice that only $f_1$-partons (or $b_2$ bosons) carries $U(1)$ charge and couples to the external $U(1)$ gauge field $A_\mu$. Here $b_\mu$ is the gauge field which enforces the constraint (\ref{constraint:u(1) spt}). The currents of $f$- and $f_1$-partons are written in terms of dual $U(1)$ gauge fields $a^f_\mu$ and $a^{f_1}_\mu$ by (\ref{duality:electron current}). We can integrate out gauge fields $a^f_\mu$ and $a^{f_1}_\mu$ in (\ref{lagrangian:nf=2}) and obtain the simplified low-energy effective theory:
\begin{eqnarray}
&\notag\mathcal{L}_{qpt}=\sum_{s=1}^2\bar{\psi}_s\gamma^\mu(\imth\partial_\mu+b_\mu)\psi_s+m\sum_{s=1}^2\bar{\psi}_s\psi_s\\
&+\frac1{g}(\epsilon^{\mu\nu\lambda}\partial_\nu b_\lambda)^2-\frac{\epsilon^{\mu\nu\lambda}}{4\pi}A_\mu\partial_\nu(A_\lambda+2b_\lambda).\label{QED3:Nf=2}
\end{eqnarray}
At critical point the parton bilinear mass $m=0$ and this is nothing but the $QED_3$ with fermion number $N_f=2$. In other words, the critical theory here is described by two flavors of Dirac fermions coupled to a \emph{noncompact} $U(1)$ gauge field $b_\mu$. $g$ is a coupling constant determined by microscopic details. The noncompactness of gauge field $b_\mu$ is guaranteed by $U(1)$ symmetry of $b_2$-bosons (and $f_1$-partons), which forbids any monopole event of $b_\mu$ gauge fields. In other words there is a $U(1)$ conservation of $b_\mu$ gauge flux.

Such a critical theory has been studied extensively in the context of algebraic spin liquids\cite{Affleck1988a,Marston1989,Wen1996,Kim1999,Rantner2001,Franz2001,Wen2002,Hermele2004,Hermele2005,Borokhov2002,Kaveh2005}. It has been shown that the long-distance, low-energy physics of (\ref{QED3:Nf=2}) is controlled by an interacting, conformally invariant fixed point\cite{Rantner2001}. At asymptotically low energy there is no free quasiparticle excitations. Notice that such a critical theory, \ie $QED_3$ with $N_f=2$ is equivalent\cite{Senthil2006} to a $O(4)$ sigma model with a topological $\Theta$-term at $\Theta=\pi$. 

In a more general case, the critical theory for continuous quantum phase transitions between a bosonic $U(1)$-SPT phase with $\sigma_{xy}=-2p$ and one with $\sigma_{xy}=-2q$, can be similarly obtained in the projective construction. The low-energy effective theory describing the phase transition is
\begin{eqnarray}
&\notag\mathcal{L}_{qpt}=\sum_{n=1}^{|p-q|}\sum_{s=1}^{2}\bar{\psi}_{n,s}\gamma^\mu(\imth\partial_\mu+b^n_\mu)\psi_{n,s}\\
&+m\sum_{n,s}\bar{\psi}_{n,s}\psi_{n,s}+\sum_{n}\frac1{g_n}(\epsilon^{\mu\nu\lambda}\partial_\nu b^n_\lambda)^2\notag\\
&-\frac{\epsilon^{\mu\nu\lambda}}{4\pi}A_\mu\partial_\nu\big[(p+q)A_\lambda+2\sum_nb^n_\lambda\big].\label{QED3:general}
\end{eqnarray}
There are $2|p-q|$ flavors of Dirac fermions $\{\psi_{n,s}|1\leq n\leq|p-q|,s=1,2\}$, coupled to $|p-q|$ different dynamic $U(1)$ gauge fields $\{b_\mu^n|1\leq n\leq|p-q|\}$ which enforce the constraints (\ref{constraint:u(1) spt}) in projective construction. The $U(1)$ conservation of $b_2$-bosons lead to the conservation of total flux of $b_\mu^n$ fields. The parton bilinear mass $m$ changes sign across this continuous phase transition. At critical point $m=0$ and (\ref{QED3:general}) is nothing but $|p-q|$ copies of $QED_3$ with $N_f=2$, in the presence of a total $U(1)_{flux}$ symmetry.\\

\begin{figure}
 \includegraphics[width=0.4\textwidth]{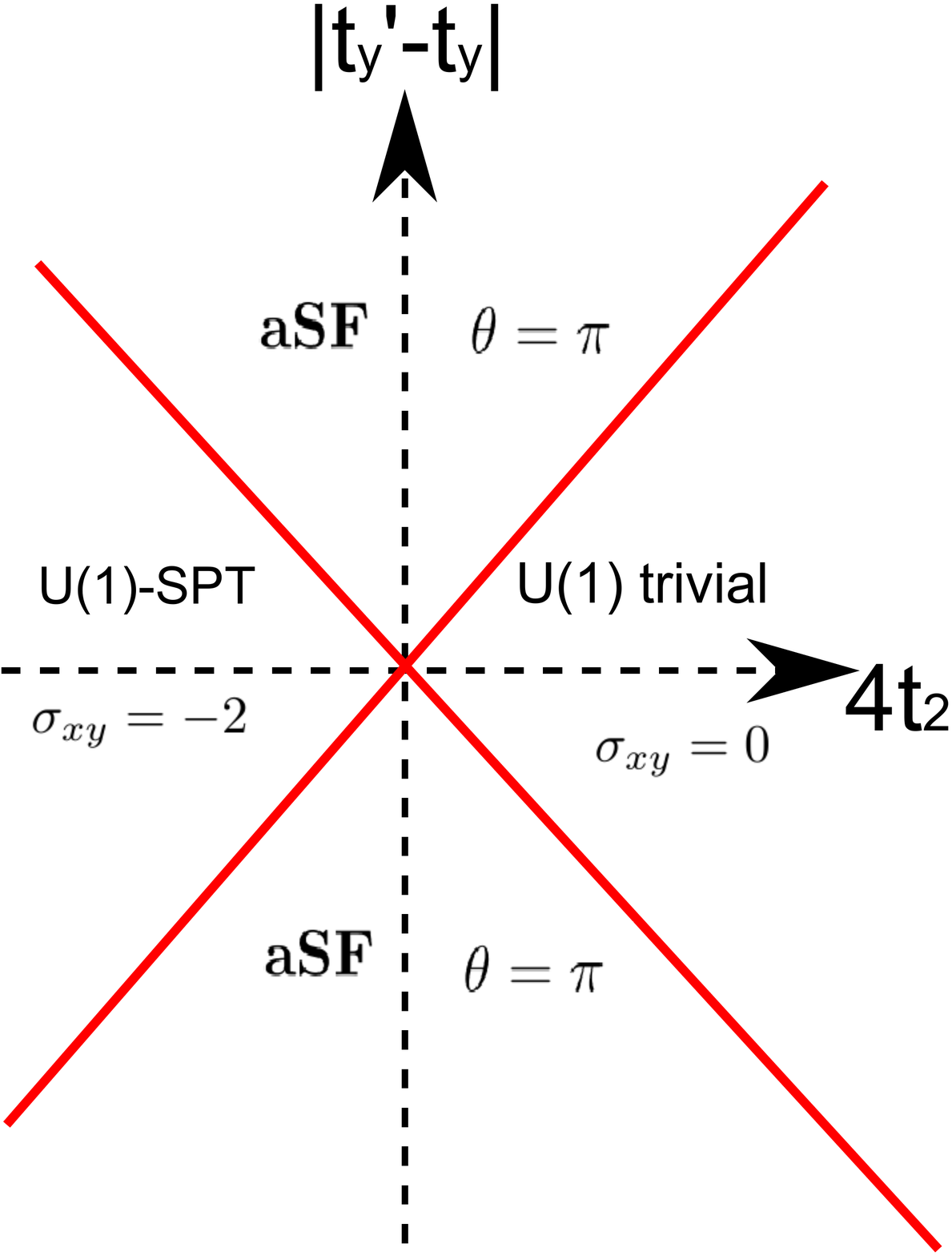}
\caption{(color online) Mean-field phase diagram of hopping ansatz (\ref{mf hopping:f1,d}) for $d_1$-partons in projective construction (\ref{parton:u(1) spt}) with $q=0$. Solid lines $|t_y^\prime-t_y|=\pm4t_2$ denote the phase boundary between trivial boson insulator with $\sigma_{xy}=0$, bosonic $U(1)$-SPT phase with $\sigma_{xy}=-2$ and anyon superfluid (aSF) with anyon statistical angle $\theta=\pi$, where continuous phase transitions happen. Chern numbers $C_f=+1$ and $C_{f_1}=-1$ are chosen for $f$- and $f_1$-partons in projective construction (\ref{parton:u(1) spt}). Notice that only when $t_y^\prime=t_y$, there is a direct continuous phase transition between bosonic $U(1)$-SPT phase and the trivial boson insulator. The effective theory describing the tricritical point at $t_y^\prime-t_y=t_2=0$ is $QED_3$ with $N_f=2$.}\label{fig:mean-field phase diagram}
\end{figure}

We want to emphasize that the above critical point, described by $QED_3$ with fermion flavor number $N_f=2|p-q|$, are the \emph{minimal} description for continuous quantum phase transitions between two bosonic $U(1)$-SPT phases with Hall conductance $\sigma_{xy}=-2p$ and $-2q$. More generally there can be more flavors of massless Dirac fermions (\ie $N_f>2|p-q|$) at the critical point. Take $p=1,q=0$ for example, the ``minimal model'' for a critical point between the trivial Mott insulator ($\sigma_{xy=0}$) and $\sigma_{xy}=-2$ bosonic $U(1)$-SPT phase is $QED_3$ with $N_f=2$ flavors of Dirac fermions, as shown in (\ref{QED3:Nf=2}). In a generic critical point between these two phases, the flavor number could be more \ie $N_f/2>1$, and all these Dirac fermions are low-energy modes at the band touching points of $d_1$-partons in (\ref{parton:u(1) spt}). Let't consider, say, $N_f=4$ in effective theory (\ref{QED3:Nf=2}) so that $s=1,2,3,4$ in the summation. This actually describes a multi-critical point, whose neighboring phases include not only superfluid, trivial Mott insulator ($\sigma_{xy}=0$) and bosonic $U(1)$-SPT phase ($\sigma_{xy}=-2$), but also AFQH states of bosons with $\sigma_{xy}=-1/2$ and $\sigma_{xy}=-3/2$.

\subsection{Intermediate phases between two different bosonic $U(1)$-SPT phases: spontaneous $U(1)$ symmetry breaking and anyon superfluid}

The continuous phase transition between different bosonic $U(1)$-SPT phases, however, is not generic and needs fine tuning to be reached. This is simply because in (\ref{QED3:general}) and (\ref{QED3:Nf=2}) the parton masses $m$ for different branches of Dirac cones always change sign simultaneously. Usually such a continuous phase transition, which requires Chern number changing by two, needs to fine tuned in the absence of extra symmetries. Certain extra symmetries, such as $C_4$ and translational symmetry of square lattice in our lattice model (\ref{mf hopping:f1,d}), can guarantee that $t_x=t_y=t_y^\prime$ and hence the mass terms change sign simultaneously at the two Dirac cones ${\bf k}=\pm(\pi/2,\pi)$. In the absence of extra symmetries, however, generically there will be intermediate phases between different $U(1)$-SPT phases , instead of a direct phase transitions.

Again let's first look at the simplest case, \ie what intermediate phase would emerge between a trivial boson insulator ($\sigma_{xy}=0$) and a bosonic $U(1)$-SPT phase with $\sigma_{xy}=-2$. Note that if one integrates out a Dirac fermion with mass $m$ coupled to $U(1)$ gauge field $a_\mu$, a Chern-Simons term $\frac{\text{Sgn}(m)}2\frac{\epsilon^{\mu\nu\lambda}}{4\pi}a_\mu\partial_\nu a_\lambda$ is obtained. Hence if the mass term changes sign for only one flavor of Dirac fermions (\ref{QED3:Nf=2}), we'll obtain the following effective theory for the intermediate phase
\begin{eqnarray}
\mathcal{L}_{aSF}=\frac1{g^\prime}(\epsilon^{\mu\nu\lambda}\partial_\nu b_\lambda)^2-\frac{\epsilon^{\mu\nu\lambda}}{4\pi}A_\mu\partial_\nu(A_\lambda+2b_\lambda).\label{Lagrangian:anyon SF}
\end{eqnarray}
This intermediate phase is gapless, featured by photon excitations of dynamical $b_\mu$ gauge fields. In fact integrating out $b_\mu$ fields one can obtain the electromagnetic response of such a state:
\begin{eqnarray}
&\notag\mathcal{L}_{response}=\tilde{g}A_\mu(\delta_{\mu,\nu}-\frac{\partial_\mu\partial_\nu}{\square})A_\nu\\
&-\frac{\epsilon^{\mu\nu\lambda}}{4\theta}A_\mu\partial_\nu A_\lambda+\cdots\label{response:anyon SF}
\end{eqnarray}
where $\cdots$ denotes higher order terms. Characterized by a superfluid response together with a \emph{quantized} Chern-Simons term, this is nothing but the electromagnetic response theory of an anyon superconductor\cite{CHEN1989,Fetter1989,Fradkin1990}, where the anyon statistical angle is $\theta=\pi$.

Anyon superconductivity was proposed as the ground state of high-$T_c$ cuprate superconductors\cite{LAUGHLIN1988a}, where it was conjectured that each hole (``holon'') doped into the antiferromagnetic parent compound has sermionic statistics\cite{Laughlin1988} $\theta=\pi/2$. Since single holon cannot condense due to its fractional statistics, they form Cooper pairs due to attractive ``statistical'' interactions\cite{Arovas1985}, which obey Bose-Einstein statistics. An anyon superconductor is the Bose-Einstein condensate of bosonic bound states formed by a multiple of anyons: it not only exhibits Meissner effect but Hall effect as well, as a manifestation of $P,T$ symmetry breaking\cite{CHEN1989,Fetter1989,Fradkin1990}.

Therefore we dub this intermediate phase an ``\emph{anyon superfluid}" (aSF). It spontaneously breaks the global $U(1)$ symmetry associated with $b_2$-boson conservation, and the gapless photon excitation in (\ref{Lagrangian:anyon SF}) corresponds to the Goldstone mode (phonon) of $U(1)$ symmetry breaking\cite{Ran2008} in aSF.

Such an intermediate phase indeed happens in lattice model (\ref{mf hopping:f1,d}) for $d_1$-parton mean-field hoppings. Once lattice translation and $C_4$ symmetry of square lattice is broken, we can choose $|t_y^\prime-t_y|>4|t_2|$ in the mean-field ansatz for $d_1$-partons, according to (\ref{chern number}) and discussions above the system immediately enters an anyon superfluid phase. Notice we always keep $C_f=+1$ and $C_{f_1}=-1$ in the process. A phase diagram of the ground state by projective construction (\ref{parton:u(1) spt})(\ref{wf:u(1) spt}), as a function of hopping parameters in mean-field anstaz (\ref{mf hopping:f1,d}) for $d_1$-partons, is shown in FIG. \ref{fig:mean-field phase diagram}. It's easy to figure out there is generically an intermediate aSF phase between $U(1)$-SPT phase with $\sigma_{xy}=2q$ and with $\sigma_{xy}=2q-2$, where anyons have statistical angle
\begin{eqnarray}
\theta=\frac{\pi}{1-2q}.\label{statistics:aSF}
\end{eqnarray}
A schematic phase diagram of interacting bosons with $U(1)$ symmetry in two dimensions, containing bosonic $U(1)$-SPT phases and aSF phases, is shown in FIG. \ref{fig:schematic phase diagram}.

The continuous phase transition between an aSF with $\theta=\frac{\pi}{1-2q}$ and a bosonic $U(1)$-SPT phase with Hall conductance $\sigma_{xy}=2q-1\pm1$, can also be easily studied based on the projective construction. The low-energy effective theory describing such a phase transition is a single Dirac fermion coupled with a $U(1)$ gauge field $b_\mu$ with a Chern-Simons term:
\begin{eqnarray}
&\notag\mathcal{L}_{aSF-SPT}=\bar{\psi}\gamma^\mu(\imth\partial_\mu+b_\mu)\psi+m\bar{\psi}\psi\\
&\mp\frac{\epsilon^{\mu\nu\lambda}}{8\pi}b_\mu\partial_\nu b_\lambda+\frac{\epsilon^{\mu\nu\lambda}}{4\pi}A_\mu\partial_\nu\Big[(2q-1)A_\lambda+2b_\lambda\Big].\label{aSF-SPT}
\end{eqnarray}
Again the gauge field $b_\mu$ is noncompact due to the $U(1)$ symmetry. Critical exponents of this theory has been calculated\cite{Chen1993} in the large-$N_f$ (flavor number of Dirac fermions) expansion.

The main difference between aSFs here and conventional superfluids is their symmetry: a conventional superfluid preserves $P,T$ symmetries while the aSF breaks them. In fact due to breaking of $U(1)$ charge conservation and associated gapless Goldstone modes, the $\theta$ angle in EM response (\ref{response:anyon SF}) is not quantized in a superfluid and can be tuned continuously. Therefore a continuous transition between bosonic $U(1)$-SPT phases and a conventional superfluid is possible.\\

At last we comment on the physical meaning of fermionic partons introduced in (\ref{parton:u(1) spt}). Putting $f$-partons in an insulating band structure with Chern number $C_f=-1$ plays the role of attaching a unit flux quantum to each boson\cite{Girvin1987,Zhang1989}. As a result bosons $b_1$ and $b_2$ together with the attached flux form composite fermions\cite{Jain1989}: they are nothing but $d_1$ and $f_1$ in (\ref{parton:u(1) spt}). Since the total boson density is bound to the $U(1)$ flux density seen by composite fermions, conservation of boson number leads to conserved total flux number. In both the trivial Mott insulator and $U(1)$-SPT phase, composite fermions $f$ and $f_1$ both form an insulator and there is a finite energy gap for all excitations in the bulk. The boson density fluctuation \ie the $U(1)$ gauge field $a_\mu$ in (\ref{QED3:Nf=2}) is also gapped. At the critical point, similar to the superfluid-Mott transition\cite{Fisher1989a}, boson density fluctuations are gapless and hence gauge field $a_\mu$ also becomes gapless.


\section{Concluding remarks}\label{CONCLUSION}

In this work we study the continuous quantum phase transitions between different bosonic $U(1)$-SPT phases in two dimensions. A projective construction is developed for $U(1)$-SPT phases of bosons in two dimensions. The projective construction not only provides a straightforward view to the quantum phase transition between different SPT phases, but also allows one to write down many-body wavefunctions for bosonic SPT phases on a lattice. Although we focus on bosonic SPT phases with $U(1)$ symmetry in this work, the projective construction can be easily generalized to other symmetry groups such as $Z_n$. We show that the continuous quantum phase transitions between two different $U(1)$-SPT phases is captured by emergent $QED_3$ with $N_f=2$. In other words, in low-energy long-wavelength limit the critical point is described by two flavors of Dirac fermions coupled to a dynamical noncompact $U(1)$ gauge field. However such a continuous phase transition is not generic and needs fine tuning in the absence of extra symmetries. We show that there is an intermediate phase between two $U(1)$-SPT phases whose Hall conductance differing by 2. This intermediate phase has the same electromagnetic response as an anyon superconductor, and is hence dubbed ``anyon superfluid". Such an anyon superfluid can be connected to a bosonic $U(1)$-SPT phase with proper $\sigma_{xy}$ by a continuous phase transition. Based on these results, a generic phase diagram of interacting bosons with $U(1)$ symmetry in two dimensions is sketched.

Upon completion of this work, we notice an independent work by Tarun Grover and Ashvin Vishwanath who studied similar problems.

\acknowledgements

YML is indebted to Ying Ran for an inspiring discussion, which eventually leads to the projective construction of bosonic SPT phases in two dimensions. YML thanks Kavli Institute for Theoretical Physics China, and Kavli Institute for Theoretical Physics for hospitality, where part of this work was finished during 2012 KITP program ``Frustrated Magnetism and Quantum Spin Liquids''. YML and DHL acknowledges the support by the DOE grant number DE-AC02-05CH11231. This research was supported in part by the National Science Foundation under Grant No. NSF PHY05-51164(YML).

%

\end{document}